 \documentstyle[epsf,twocolumn,prb,aps]{revtex}

\begin{document}
\draft

 \twocolumn[\hsize\textwidth\columnwidth\hsize\csname @twocolumnfalse\endcsname

\title{Metal-insulator transition in the one-dimensional Holstein model at
half filling}
\author{Eric Jeckelmann \cite{byline}, Chunli Zhang and Steven R. White}
\address{Department of Physics and Astronomy, University of California,
Irvine, CA 92697.
}
\date{\today}
\maketitle
\begin{abstract}
We study the one-dimensional Holstein model with spin-1/2 electrons at 
half-filling.  
Ground state properties are calculated for long chains 
with great accuracy using the density matrix renormalization group method
and extrapolated to the thermodynamic limit.
We show that for small electron-phonon coupling or large phonon frequency,
the insulating Peierls ground state predicted by mean-field theory is 
destroyed by quantum lattice fluctuations and that
the system remains in a metallic  phase with a non-degenerate
ground state and power-law electronic and phononic correlations. 
When the electron-phonon coupling  becomes large or the phonon frequency 
small,
the system undergoes a transition to an insulating Peierls phase with a 
two-fold degenerate ground state, long-range charge-density-wave order, 
a dimerized lattice structure, and
a gap in the electronic excitation spectrum. 
\end{abstract}


 ]

\narrowtext

A long time ago Peierls \cite{Peierls} suggested that a one-dimensional
metal should exhibit an instability against a periodic lattice distortion
of wave vector equal to twice the Fermi wave vector.
Although this distortion increases the lattice elastic energy, it opens
a gap in the electronic spectrum at the Fermi surface, lowering the 
electronic energy. 
Thus, the Peierls insulating state can be energetically favored over the
metallic state.
A wide range of quasi-one-dimensional materials, such as MX chains,
charge-density-wave (CDW) compounds, conjugated polymers and charge-transfer
salts, \cite{1dmaterials} have electronic properties that are dominated 
or at least affected by the Peierls instability.
These systems are often modeled by the one-dimensional Holstein model,
\cite{Holstein} the Su-Schrieffer-Heeger model \cite{ssh80} or various
spin-Peierls~\cite{sp} models.

The Peierls instability is well understood in the static lattice (adiabatic)
limit and within mean-field theory.
An interesting and still controversial question is how the Peierls ground 
state is modified when quantum lattice fluctuations are taken into account.
These quantum lattice fluctuations could have an important effect in most 
quasi-one-dimensional materials with a Peierls ground state because the 
lattice zero-point motion is often comparable to the amplitude of the 
Peierls distortion.~\cite{kw92}
Thus, this question has motivated several studies of quantum lattice 
fluctuation effects in the Holstein 
\cite{hf83,cb84,zfa89,whs95,mkhm96,bmkh98,wf98,za98}, 
Su-Schrieffer-Heeger \cite{su82,fh83,tak92,qcra93,zhe94} and 
spin-Peierls \cite{cm96,ssc97,zhe97,apsa98,wel98,bur98,wei99,san99} models.
In spinless fermion models and spin-Peierls models these studies 
have shown that the transition to a Peierls state occurs only when
the electron-phonon coupling exceeds a finite critical value or when the 
phonon frequency drops below some finite threshold value.
Thus, in these systems quantum lattice fluctuations destroy the 
Peierls instability for small electron-phonon coupling or large phonon 
frequency.
In more realistic models with spin-1/2 electrons, however, 
previous studies~\cite{hf83,zfa89,fh83,tak92,zhe94} have generally concluded 
that the ground state is a Peierls state for any finite 
electron-phonon coupling at finite phonon frequency, in qualitative 
agreement with mean-field theory. 

Here we consider the one-dimensional Holstein model with spin-1/2
electrons at half-filling.
This model describes electrons coupled to dispersionless phonons, 
represented by local oscillators.
It has as Hamiltonian
\begin{eqnarray}
H =  
\frac{1}{2M} \, \sum_i p^2_i \, + \, \frac{K}{2} \sum_i q^2_i
- \alpha \, \sum_i  q_i ( n_i - 1 ) \nonumber \\
- t \sum_{i\sigma} \left (c^\dag_{i\sigma}
c_{i+1\sigma} + c^\dag_{i+1\sigma} c_{i\sigma} \right )  \, ,
\label{eq:ham1}
\end{eqnarray}
where $q_i$ and $p_i$ are the position and momentum operators
for a phonon mode at site $i$,
$c^\dag_{i,\sigma}$ and $c_{i,\sigma}$ are creation and 
annihilation operators for an electron of spin 
$\sigma$ on site $i$, 
and $n_i = c^\dag_{i,\uparrow} c_{i,\uparrow}
+c^\dag_{i,\downarrow} c_{i,\downarrow}$.
The half-filled band case corresponds to a density of one electron
per site.
At first sight, there are four parameters in this model: the oscillator
mass $M$ and spring constant $K$,
the electron-phonon coupling constant $\alpha$,
and the electron hopping integral $t$.
However, if phonon creation and annihilation operators are denoted by 
$b^\dag_i$ and $b_i$, respectively,
the Holstein Hamiltonian can be written (up to a constant term) 
\begin{eqnarray}
H =  \omega \, \sum_i b^\dag_i b_i
- \gamma \, \sum_i \left (b^\dag_i + b_i \right ) (n_i - 1) \nonumber \\ 
- t \sum_{i\sigma} \left (c^\dag_{i+1\sigma} c_{i\sigma} 
+ c^\dag_{i\sigma} c_{i+1\sigma} \right )  \, ,
\label{eq:ham2}
\end{eqnarray}
where the phonon frequency is given by $\omega^2 = K/M$ (we set $\hbar = 1$) 
and a new electron-phonon constant is defined by $\gamma = \alpha \, a$
with the range of zero-point phonon position fluctuations 
given by $a^2 = \omega/2K$.
We can set the parameters $t$ and $a$ equal to 1 by 
redefining the overall energy scale and the units of phonon displacements.
Thus, the properties of the Holstein Hamiltonian (\ref{eq:ham2}) 
depends only on the two interaction parameters $\omega$ and $\gamma$.

Mean-field theory predicts that the ground state of this model
is a Peierls state for any non-zero electron-phonon coupling and
$\omega < \infty$.
Early works based on strong-coupling perturbation theory
and quantum Monte-Carlo simulations,~\cite{hf83} as well as  
variational calculations~\cite{zfa89} seemed to support this point of view. 
However, the quantum Monte-Carlo results were limited to small systems
(up to 16 sites) and their interpretation relied on a questionable
finite-size-scaling analysis.
The strong-coupling perturbation theory is based on the formation
of small bipolarons in the $\gamma/\omega \rightarrow \infty$ limit,
but it has been argued that, as the coupling $\gamma$ decreases,
the bipolaron size becomes large and the strong-coupling picture breaks
down~\cite{aar92}.  
On the other hand, a functional integral calculation suggests that the 
transition occurs at finite electron-phonon coupling\cite{whs95}, but the
accuracy of this approach is hard to estimate.
Moreover, the static and dynamical properties of small clusters
(up to six sites) show that there is a sharp crossover 
at a finite electron-phonon coupling   
from a quasi-free electron ground state to an ordered bipolaronic 
ground state, which can be seen as a precursor to the Peierls ground state
of the infinite system.~\cite{2sites,zjw98,chunli}

In this paper we discuss the ground state properties of
the Holstein model of spin-1/2 electrons in the thermodynamic limit.
We demonstrate that quantum lattice fluctuations suppress the Peierls 
instability for small electron-phonon coupling or large 
phonon frequency. 
In this regime the ground state is unique, gapless and shows
only power-law correlations between electron position and between
phonon displacements. 
This ground state is similar to the ground state of the 
non-interacting system ($\gamma = 0$).
When the electron-phonon coupling  becomes large or the phonon frequency 
becomes small the system undergoes a transition to an insulating Peierls 
phase, which is qualitatively described by mean-field theory.
In this regime the ground state is doubly degenerate, and
there is a gap in the electronic spectrum, long-range
CDW order and a dimerized lattice structure.

Our results are based on density matrix renormalization group (DMRG) 
calculations.\cite{dmrg}
DMRG is as accurate as exact diagonalization on small systems but
can be applied to much larger systems while maintaining very good precision.
It has already been applied successfully to the study of the Peierls
instability in quantum lattices with       
spinless fermion or spin degrees of freedom.~\cite{cm96,bmkh98,bur98}
There have not been any application of DMRG to models of spin-1/2 electrons
coupled to phonons yet, because these systems are significantly harder
to deal with due to the additional degrees of freedom and the larger
amplitude of phonon displacements.
For this work we have used an improved DMRG method for systems with boson
degrees of freedom, which has been described in a 
previous work.~\cite{bosondmrg}
With this approach both the error due to the necessary truncation of the 
phonon Hilbert space and the DMRG truncation error can be kept negligible.
The accuracy of this new DMRG technique has been demonstrated by comparison 
with many numerical and analytical methods for the polaron problem
(a single electron) in the one-dimensional Holstein 
model.~\cite{bosondmrg,rbl98,tru98}
The maximum number of density matrix eigenstates $m$ used
in our calculations is 600, giving truncation errors from $10^{-7}$ to
$10^{-11}$ depending on the system size and parameters. 
The error in the ground state energy is estimated to be smaller than 
$10^{-5} t$.
The actual number of phonon states kept for each local oscillator ranges
from 8 to 32 depending on the electron-phonon coupling strength.
We have studied open chains with an even number $N$ of sites (up to 100)
and extrapolate results to the thermodynamic limit.
Open boundary conditions are used because the 
DMRG method usually performs much
better in this case than for periodic boundary conditions.

In previous studies of the Peierls instability in the Holstein model
the ground state symmetry was explicitly 
broken as in the mean-field and adiabatic 
approximations.~\cite{hf83,zfa89,whs95} 
Thus, the Peierls ground state was revealed by 
a lattice distortion (dimerization) 
\begin{equation}
\langle q_i \rangle = (-1)^i m_p
\label{eq:dim}
\end{equation}
and a CDW
\begin{equation}
\langle n_i \rangle = 1 + (-1)^i m_e
\label{eq:cdw} 
\end{equation} 
with $m_e, m_p \neq 0$,
where $\langle \hat{O} \rangle$ means the ground state expectation
value of operator $\hat{O}$, and $m_p$ and $m_e$ are the phonon
and electronic order parameter, respectively.~\cite{hf83}
For $m_e, m_p \neq 0$, the ground state was two-fold degenerate
[this degeneracy corresponds to the two possible phases of the
oscillations (\ref{eq:dim}) and (\ref{eq:cdw})].
Note that, as all eigenstates of the Holstein Hamiltonian satisfy
\begin{equation}
\langle q_i \rangle = \frac{\alpha}{K} \, ( \, \langle n_i \rangle 
- 1 \, ) \; ,
\label{eq:self}
\end{equation}
the order parameters are related by 
\begin{equation}
m_p = \frac{\alpha}{K} m_e .
\label{eq:self2}
\end{equation}

Our DMRG method gives an excellent approximation to
the exact ground state of the Holstein
model on a lattice of finite size. 
It is known exactly that the ground state of the half-filled
Holstein model on a finite lattice is unique for $\omega \neq 0$,
implying that there is no degenerate broken symmetry ground state
at any finite electron-phonon coupling or non-zero phonon 
frequency.\cite{fl95}
Instead, there is a quasi-degeneracy of the ground state when the
electron-phonon coupling exceeds a finite critical value~\cite{chunli} 
(this point will discussed in more detail later).
Therefore, we always find $\langle q_i \rangle =  0$ and
$\langle n_i \rangle = 1$ in our calculations.
This property follows directly from the uniqueness of the ground state
and the electron-hole symmetry, i.e., the invariance of the
Hamiltonian (\ref{eq:ham1}) under the transformation
\begin{eqnarray}
c^{\dagger}_{i\sigma}  \rightarrow (-1)^i c_{i\sigma} \; , \; \;
q_i   \rightarrow - q_i .
\end{eqnarray}
To observe the consequences of the Peierls instability we have 
to look at correlation functions. The most important ones for a 
Peierls state are the staggered charge density correlation function
\begin{equation}
C_n(m) = (-1)^m (\langle n_i n_{i+m} \rangle -1 )
\end{equation}

\begin{figure}
\epsfxsize=3.175 in\centerline{\epsffile{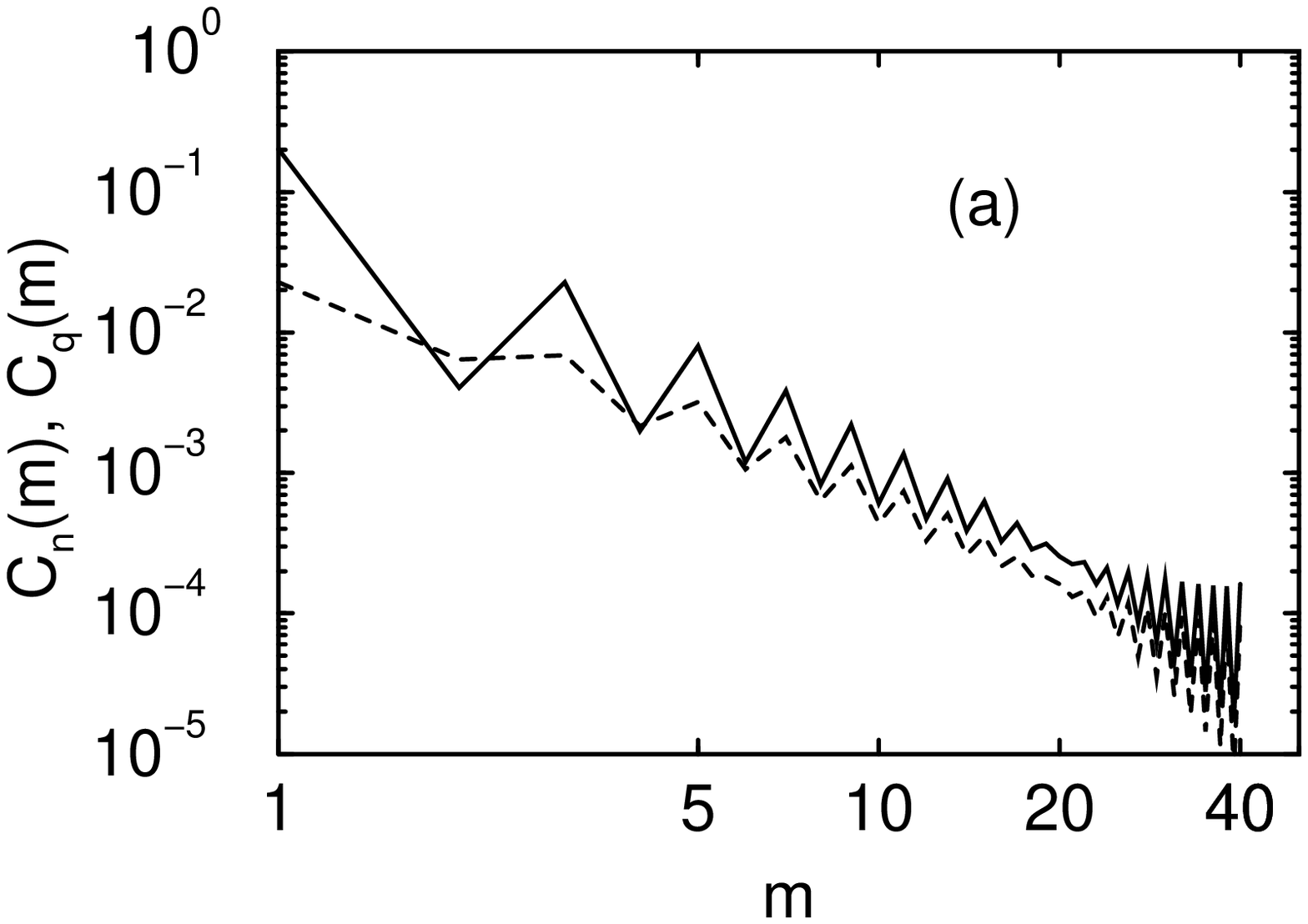}}
\end{figure}
\vspace{-1.3cm}
\begin{figure}
\epsfxsize=3.175 in\centerline{\epsffile{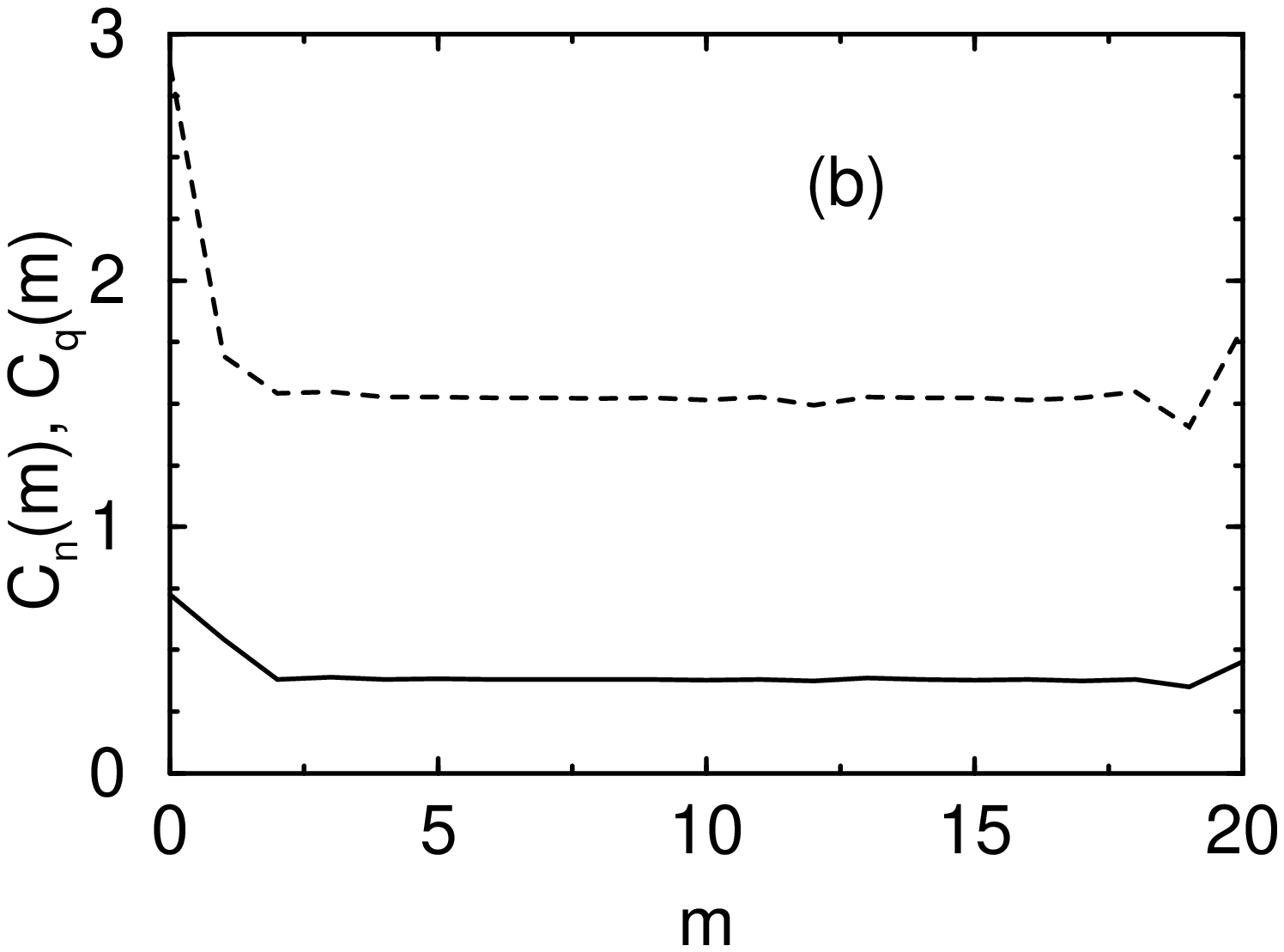}}
\caption{Staggered charge density correlation function $C_n(m)$
(solid line) and staggered
phonon displacement correlation function $C_q(m)$
(dashed line) in the metallic phase for $\gamma = 0.4$
(a) and in the Peierls phase for $\gamma = 1$ (b).
The distance $m$ is calculated from the middle
of an open chain of 80 sites (a) and 40 sites (b), respectively.
In both cases $\omega = 1$.}
\label{fig:correl}
\end{figure}

\noindent and the staggered phonon displacement correlation function
\begin{equation}
C_q(m) = (-1)^m \langle q_i q_{i+m} \rangle .
\end{equation}
We have found that, for small electron-phonon coupling $\gamma$
or large phonon frequency $\omega$, both correlation functions decrease 
as a power-law $m^{-\beta}$ with $2 \geq \beta > 0$
as a function of the distance $m$.
An example is shown in Fig.~\ref{fig:correl}(a).
As the electron-phonon coupling increases or the phonon frequency
decreases, the exponent $\beta$ becomes smaller.
For sufficiently large electron-phonon coupling or small phonon frequency
the behavior of both correlation functions is completely different.
As seen in Fig.~\ref{fig:correl}(b), in this case both functions tend  
to finite values at large distances, showing the existence of long-range order.

It is not always possible to determine the presence or absence of 
long-range order in the thermodynamic limit
from the correlation functions of a finite chain.
A better approach is to compute the electronic and phononic static 
staggered susceptibilities  defined as
\begin{equation}
\chi_e = \frac{1}{N} \sum_{m} C_n(m) 
\label{eq:chi_e}
\end{equation}

\begin{figure}
\epsfxsize=3.175 in\centerline{\epsffile{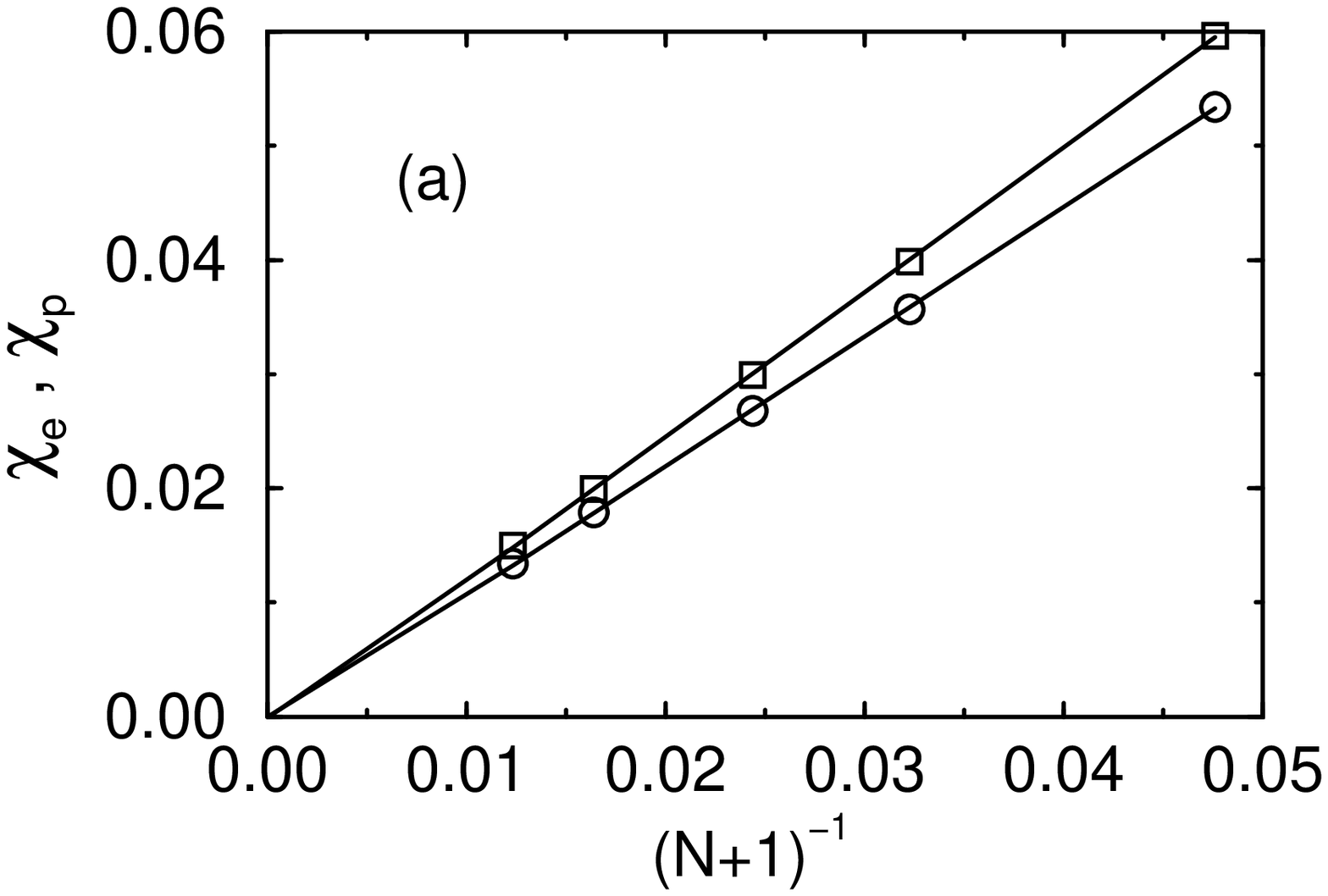}}
\end{figure}
\vspace{-1.3cm}
\begin{figure}
\epsfxsize=3.175 in\centerline{\epsffile{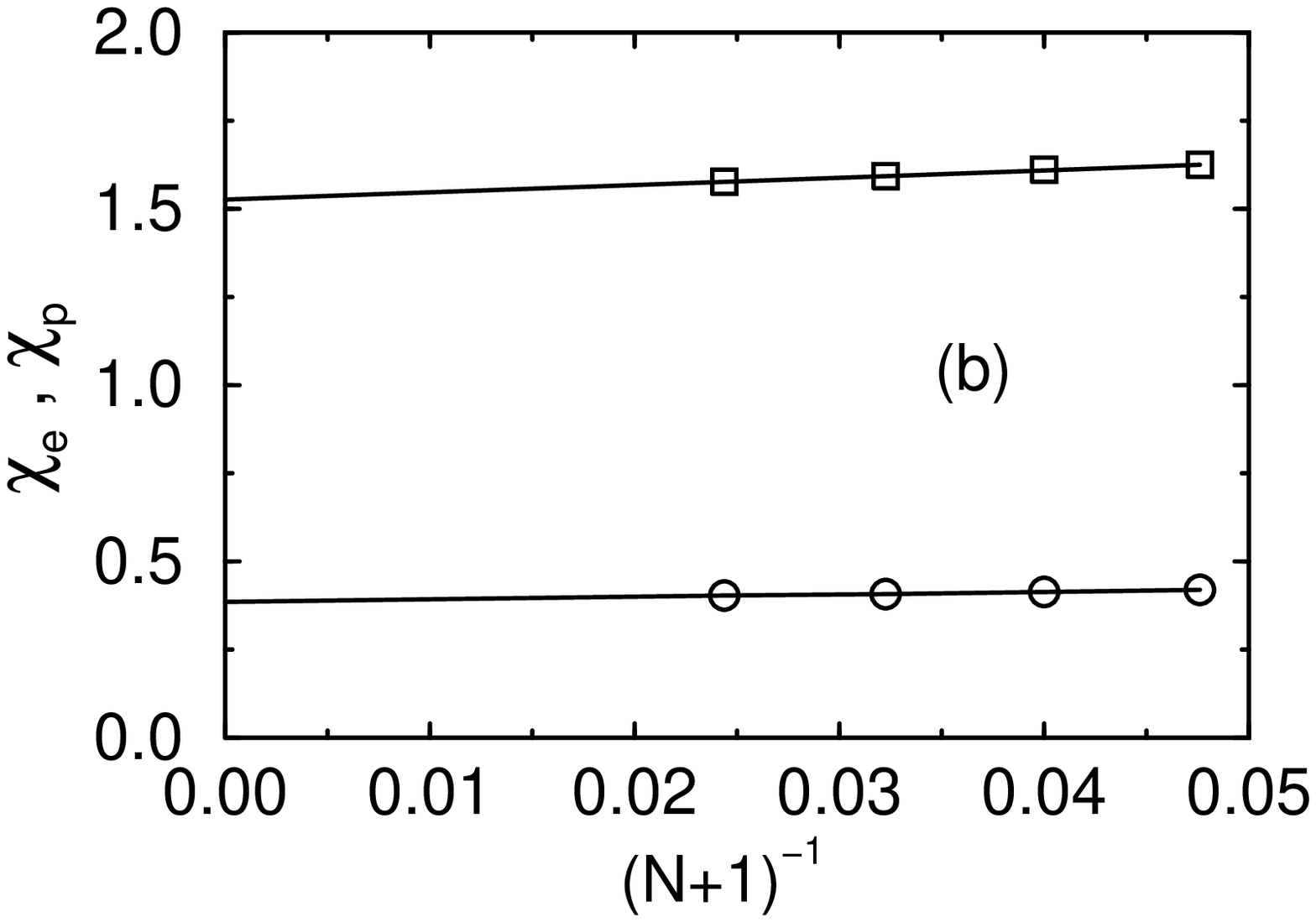}}
\caption{Electronic (circle) and phononic (square) staggered static
susceptibilities as a function of the inverse chain length
in the metallic phase for $\gamma = 0.4$ (a) and in the Peierls phase
for $\gamma = 1$ (b).
In both cases $\omega = 1$.
Solid lines are linear fits.}
\label{fig:chi}
\end{figure}

\noindent and
\begin{equation}
\chi_p = \frac{1}{N} \sum_{m} C_q(m) \, ,
\label{eq:chi_p}
\end{equation}
respectively.
It is clear that both $\chi_e$ and $\chi_p$ vanish in the thermodynamic
limit if there is no long-range order.
For instance, both susceptibilities vanish as $1/N$ in the non-interacting
limit ($\gamma = 0$).
In Fig.~\ref{fig:chi}(a) we show both $\chi_e$ and $\chi_p$ as a function
of the inverse chain length for a weak electron-phonon coupling.
Both quantities clearly tend to zero in an infinite chain. 
Thus, we conclude that there is no long-range CDW order nor lattice
distortion in the ground state of the Holstein model
for the parameters ($\gamma = 0.4, \omega = 1$) used in this example. 
On the other hand, it is clear that  $\chi_e$ and $\chi_p$  remain finite for 
$N\rightarrow \infty$ if there is long-range CDW order or a lattice
dimerization, respectively.
For instance, in the mean-field approximation, one finds
\begin{equation}
\chi_e = m_e^2  \; \; \; , \; \; \;  \chi_p = m_p^2 \; .
\label{eq:chi_mf}
\end{equation}
Figure ~\ref{fig:chi}(b) shows $\chi_e$ and $\chi_p$ as a function of the 
inverse system size for a relatively strong electron-phonon coupling. 
In this case, both susceptibilities remain finite for $N \rightarrow \infty$
and thus, reveals the presence of a Peierls state
with long-range CDW order and lattice dimerization for
the parameters considered in this example ($\gamma = 1, \omega = 1$).

Using Eqs. (\ref{eq:self2}) and (\ref{eq:chi_mf}), one sees that
\begin{equation}
\sqrt{\chi_p} = \frac{\alpha}{K} \sqrt{\chi_e}
\label{eq:rel}
\end{equation}
in the mean-field approximation.
It is possible to demonstrate that this relation holds for 
the exact ground state in several special cases, such
as the adiabatic limit ($\omega \rightarrow 0$) and
the anti-adiabatic limit ($\omega \rightarrow \infty$).
Although we can not prove the validity of (\ref{eq:rel}) for the 
general case, our numerical results show that it is 
always satisfied (within numerical errors) in an infinite system.    
This simply means that lattice dimerization and CDW are two inseparable
features of the Peierls ground state.
Therefore, we define a unique order parameter $\Delta$ as
\begin{equation}
\Delta = \alpha \sqrt{\chi_p} \approx \frac{\alpha^2}{K} \sqrt{\chi_e} \; ,
\label{eq:delta}
\end{equation}
where $\chi_p$ and $\chi_e$ are the infinite system extrapolation
of the ground state susceptibilities (\ref{eq:chi_e}) and 
(\ref{eq:chi_p}) calculated from DMRG simulations.
If the ground state of the Holstein model is a Peierls state, one
has $\Delta > 0$, and otherwise $\Delta = 0$.
Obviously, this definition of $\Delta$ is just a generalization of the 
usual gap parameter of mean-field theory $\Delta_{MF}$, which is related
to the other mean-field order parameters $m_e$ and $m_p$ by
\begin{equation}
\Delta_{MF} = \alpha | m_p | = \frac{\alpha^2}{K} | m_e|\; .
\label{eq:mf_gap}
\end{equation}

In the mean-field approximation the Peierls distortion
opens a gap $2\Delta_{MF}$ in the electronic spectrum.
It is sometimes assumed that this relation between Peierls gap and
order parameters remains valid when quantum 
lattice fluctuations are taken into account.~\cite{whs95}
In such a case the exact Peierls gap would simply be given by $2\Delta$.
However, it is likely that the Peierls gap is more reduced by the 
quantum lattice fluctuations than the dimerization or CDW amplitude~\cite{kw92}
and becomes smaller than the value $2\Delta$ obtained from
(\ref{eq:delta}).
Unfortunately, calculating the optical gap of the Holstein model
with a DMRG method is not possible yet.~\cite{till}
To find how the appearance of the Peierls ground state correlates
with a gap in the infinite system  
we have calculated the charge gaps
\begin{equation}
E_{g1} = 2 ( E_0(1)-E_0(0) )
\end{equation}
and 
\begin{equation}
E_{g2} = E_0(2)-E_0(0)  \, ,
\end{equation}
where $E_0(x)$ is the DMRG ground state energy with $x$ electrons added 
to ($x > 0$) or removed from ($x < 0$) the half-filled band.
In these definitions we implicitly use the electron-hole symmetry
of the model at half filling, which implies that $E_0(-x) = E_0(x)$.
It should be noted that with these definitions the charge gaps
incorporate lattice 

\begin{figure}
\epsfxsize=3.175 in\centerline{\epsffile{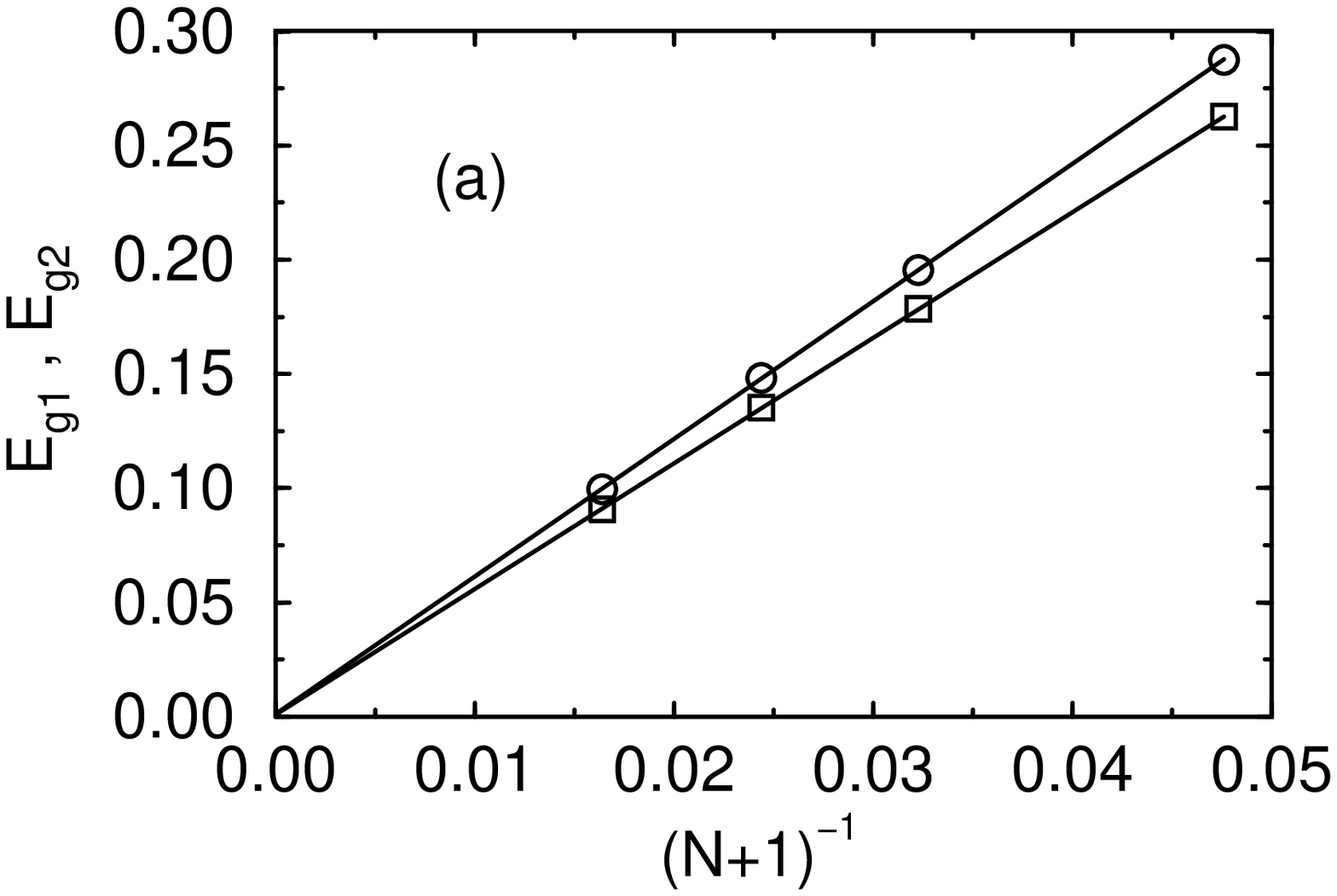}}
\end{figure}
\vspace{-1.3cm}
\begin{figure}
\epsfxsize=3.175 in\centerline{\epsffile{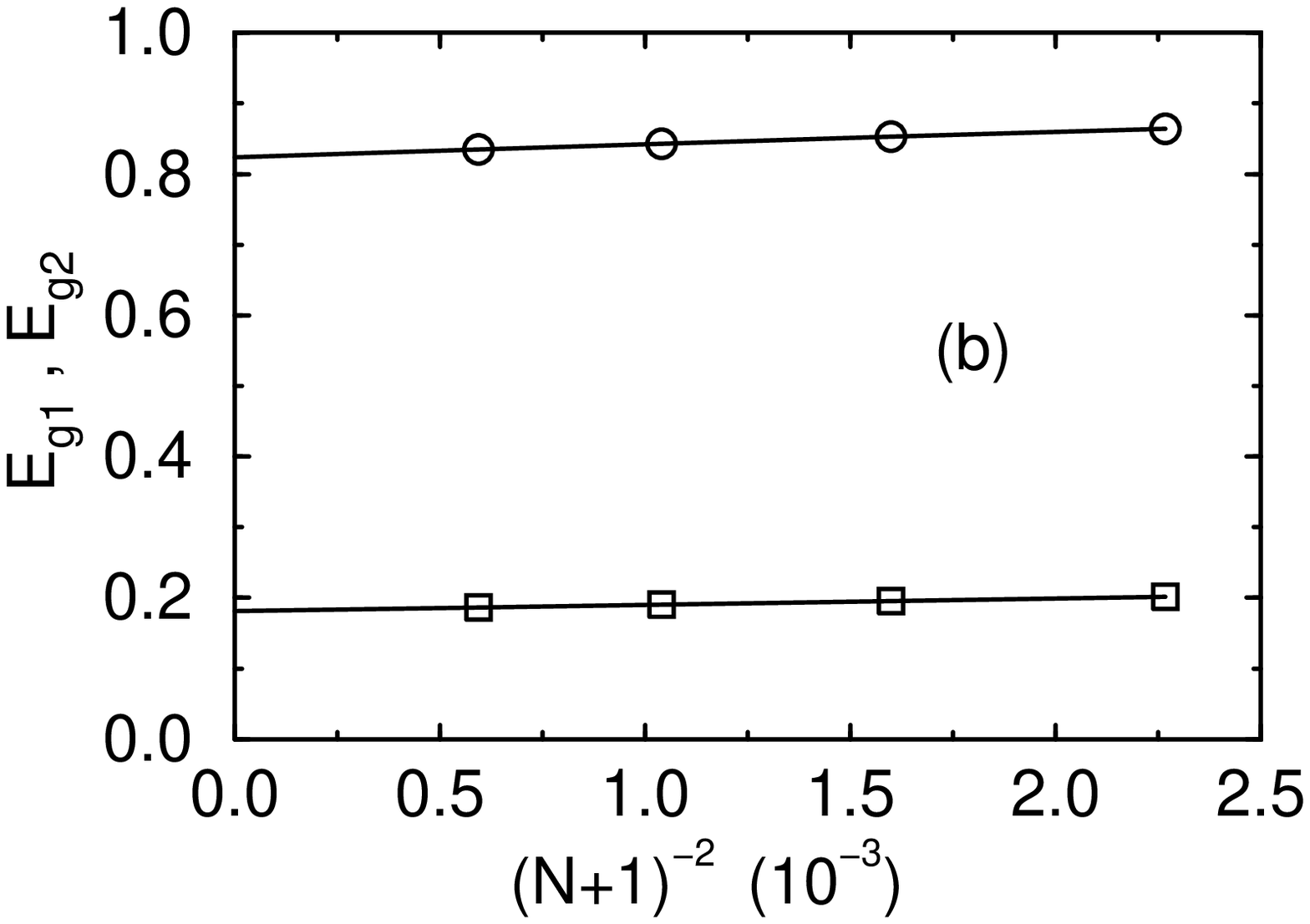}}
\caption{Gaps $E_{g1}$ (circle) and $E_{g2}$ (square) vs. the
inverse chain length in the metallic phase for $\gamma = 0.4$ (a)
and vs. the square of the inverse chain length in the Peierls phase
for $\gamma = 1$ (b).
In both cases $\omega = 1$.
Solid lines are linear fits.}
\label{fig:gaps}
\end{figure}

\noindent relaxation effects occurring
when the band filling is modified.
Therefore, $E_{g1}$ and $E_{g2}$ are not always equal to the optical
gap of the system.
$E_{g1}$ can be interpreted as the energy required to create 
a quasi-particle excitation made of an electron 
dressed by phonons.
Similarly, $2E_{g2}$ represents the energy required to create 
a quasi-particle excitation which is a bound pair
of electrons dressed by phonons, when such electron binding occurs
($E_{g2} < E_{g1}$).
Otherwise, one expects $E_{g2} \approx E_{g1}$.
Figures~\ref{fig:gaps}(a) and (b) show both gaps for several system sizes.
If there is no long-range order ($\Delta = 0$) 
we find that the gaps extrapolate to zero in the limit 
$N\rightarrow \infty$ [Fig.~\ref{fig:gaps}(a)].
Therefore, we think that in this regime the system is still a 
metal, as in the non-interacting case ($\gamma = 0 $).
However, if the ground state of the infinite system is a Peierls state
($\Delta > 0$),
we find that both gaps extrapolate to a non-zero value
in the thermodynamic limit [Fig.~\ref{fig:gaps}(b)].
For $\gamma = 1$ and $\omega = 1$, $E_{g1} = 0.82$ and
$E_{g2} = 0.18$, which are much smaller than the value that one would
anticipate from the amplitude of the Peierls distortion
$2\Delta = 2.5$.
For comparison, the mean-field result for the same parameters
is $2\Delta_{MF} = 3.1$.
This confirms that the quantum lattice fluctuations have a much 
stronger effect on the Peierls gap than on the amplitude of the
Peierls distortion.~\cite{kw92}  
Nevertheless, we have never found that either $E_{g1}$ or $E_{g2}$
vanishes for $N \rightarrow \infty$ in the Peierls ground state. 
In small clusters, a sharp 

\begin{figure}
\epsfxsize=3.175 in\centerline{\epsffile{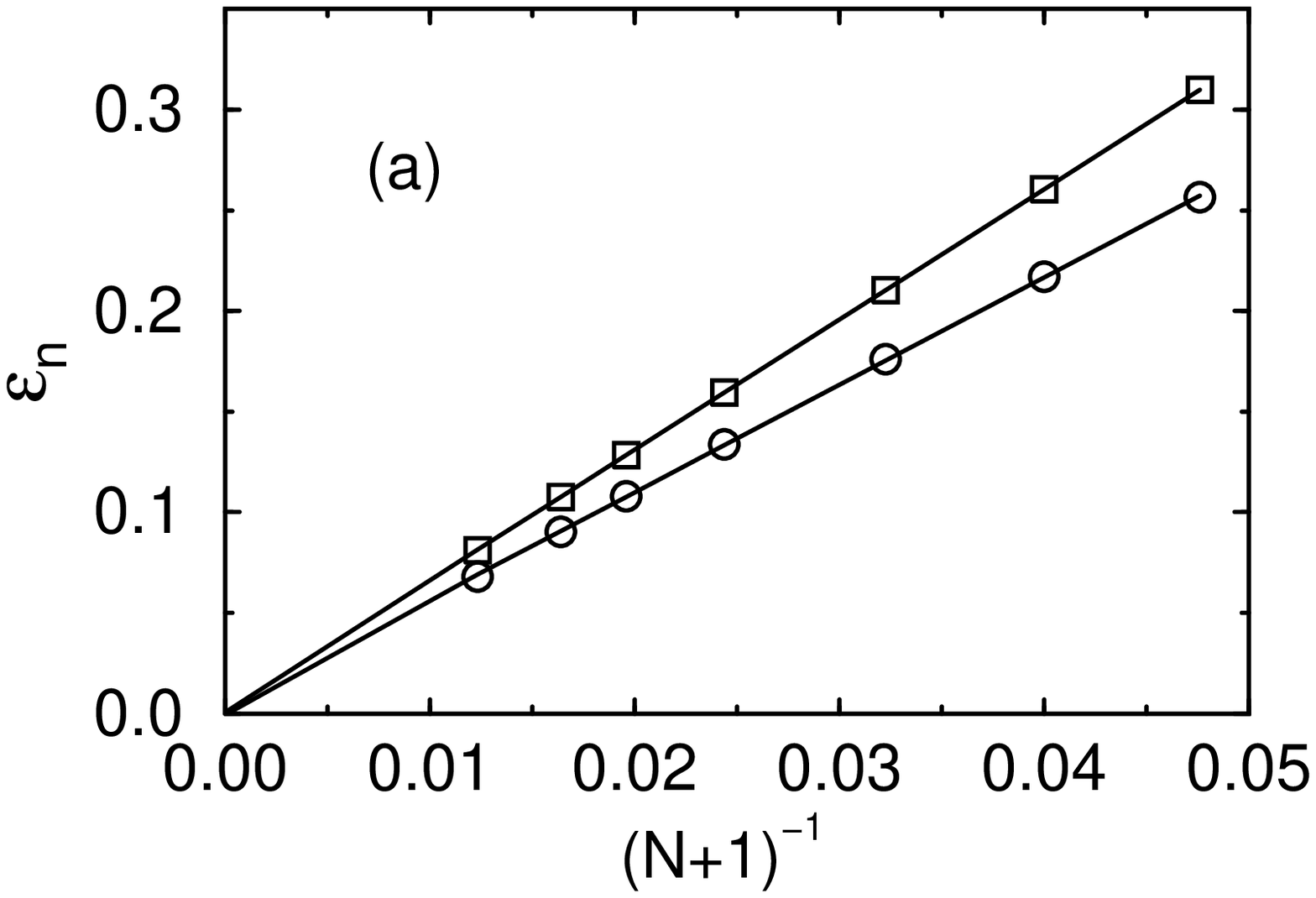}}
\end{figure}
\vspace{-1.3cm}
\begin{figure}
\epsfxsize=3.175 in\centerline{\epsffile{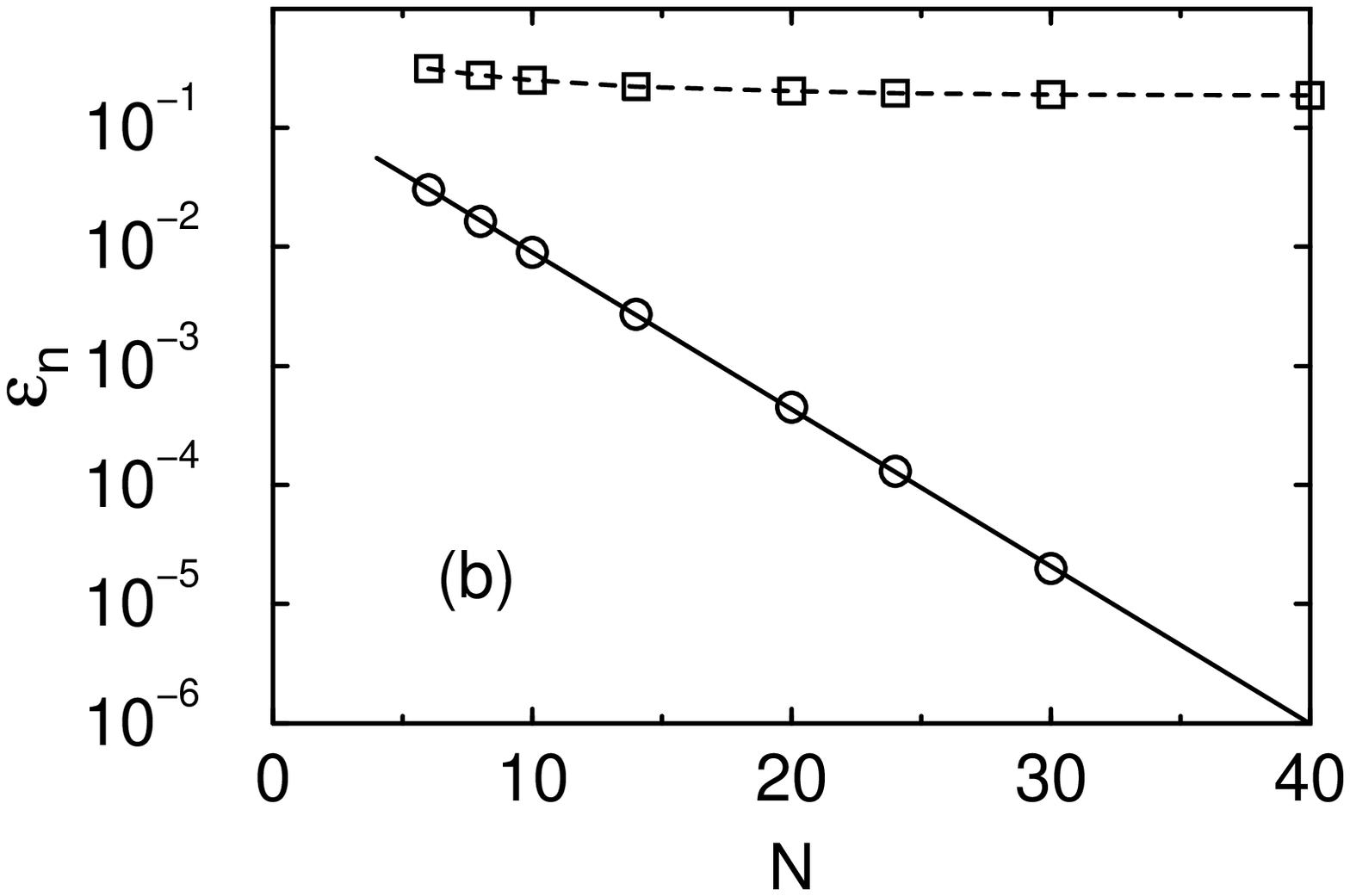}}
\caption{Lowest excitation energy $\varepsilon_1$
(circle) and second lowest excitation energy $\varepsilon_2$
(square) as a function of the system size $N$
in the metallic phase for $\gamma = 0.4$ (a)
and in the Peierls phase for $\gamma = 1$ (b).
In both cases $\omega = 1$.
Solid lines are linear fits.}
\label{fig:energies}
\end{figure}

\noindent drop of the Drude weight occurs simultaneously
with the crossover to the ordered bipolaronic ground state.~\cite{chunli}
Therefore, the opening of the electronic gap always seems to accompany
the appearance of long-range order in the ground state and
we conclude that a Peierls ground state is always an insulator. 

We have also analyzed the scaling of the lowest excitation energies
$\varepsilon_n = E_n - E_0$ with the system size,
where $E_n$ is the energy of the $n$-th lowest
eigenstate of the Hamiltonian (\ref{eq:ham2})
at half filling.
In the phase without long-range order we have found that
the $\varepsilon_n$ decrease as a power-law for increasing system size
and vanish in the thermodynamic limit, as seen in Fig.~\ref{fig:energies}(a).
These results confirm that in this case the infinite system has a unique 
ground state but is gapless; 
there is a continuous band of excitations starting 
from the ground state, as expected for a metal.
In the Peierls phase, 
the energy difference $\varepsilon_1$ between the ground state and 
the first excited state is very small even in small chains and the 
other excited states have a much higher energy.
Thus, the ground state appears almost degenerate in finite systems. 
Moreover, we observe completely different scalings for
the $\varepsilon_n$. 
Figure~\ref{fig:energies}(b) shows that $\varepsilon_1$
decreases exponentially with increasing system size, while
the energy differences between the two lowest eigenstates 
and the higher excited states 

\begin{figure}
\epsfxsize=3.175 in\centerline{\epsffile{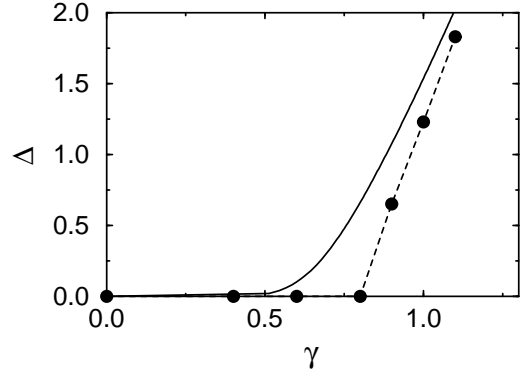}}
\caption{Order parameter $\Delta$ (circle)
as a function of the electron-phonon coupling $\gamma$ for $\omega=1$.
The solid line is the mean-field result $\Delta_{MF}$.}
\label{fig:delta}
\end{figure}

\noindent remain finite in the thermodynamic limit.
This shows that the ground state of the Peierls phase
is two-fold degenerate in the thermodynamic limit. 
We have also checked that the order parameter $\Delta$ calculated for
the first excited state tends to the same finite value as for the ground
state in the thermodynamic limit.
Therefore, both states are Peierls states 
with long-range CDW order and lattice dimerization, 
in qualitative agreement with mean-field predictions. 
The gap between the degenerate ground state
and the other eigenstates also confirms the insulating nature of 
the system in the Peierls phase.

\begin{figure}
\epsfxsize=3.175 in\centerline{\epsffile{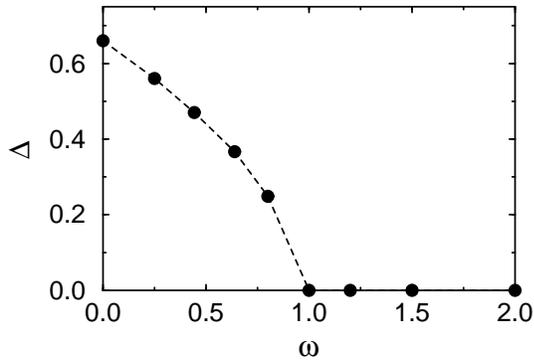}}
\caption{Order parameter $\Delta$ 
as a function of the phonon frequency $\omega$ for 
$\lambda = \gamma^2/\omega = 0.64$.
For $\omega = 0$ we show the exact adiabatic result.}
\label{fig:deltaomega}
\end{figure}

Our results demonstrate that the ground state of the one-dimensional
Holstein model for spin-1/2 electrons at half filling can be either
a metallic state or an insulating Peierls state depending on the
interaction parameters $\gamma$ and $\omega$.
The system undergoes a quantum phase transition between the 
metallic phase and the Peierls insulating phase at finite critical values 
$\gamma_c$ and $\omega_c$.
In this aspect, the Holstein model for spin-1/2 electrons is similar to
spin-Peierls and spinless fermion models.
Unfortunately, DMRG simulations become less accurate and harder to carry 
out in the vicinity of the transition while, at the same time,
the finite-size-scaling analysis requires more accurate results and 
larger system sizes.
Therefore, determining the critical values $\gamma_c$ and $\omega_c$ 
for which this metal-insulator transition occurs
demands a substantial amount of computer time 
and we have not attempted to draw a phase diagram.
Nevertheless, we can show the evolution of the order parameter
$\Delta$ as a function of the electron-phonon $\gamma$ for $\omega = 1$
in Fig.~\ref{fig:delta}.
We see that the transition to the Peierls state occurs around
$\gamma = 0.8$.
This is in good agreement with calculations based on a functional
integral approach,~\cite{whs95} which predicts $\gamma_c \approx 1$ 
for a slightly larger phonon frequency $\omega =1.1$.
As the adiabatic and anti-adiabatic limits are usually investigated
for finite values of the electron-phonon coupling constant
$\lambda = \alpha^2/2K$ ($=\gamma^2/\omega$ with our choice of units),
we show $\Delta$ as a function of the phonon frequency $\omega$
for a fixed value $\lambda=0.64$ in Fig.~\ref{fig:deltaomega}.
One can see that our results converge to the exact adiabatic result
for small $\omega$ and 

\noindent that the transition from the Peierls
phase to the metallic phase occurs around $\omega=1$. 

In summary, we have studied the ground state properties
of the one-dimensional
Holstein model for spin-1/2 electrons at half filling using DMRG.
We have shown that this system undergoes a transition
from a metallic phase to an insulating Peierls phase 
at finite values of the electron-phonon coupling and of 
the phonon frequency.

\acknowledgments
We thank S. Moukouri and I. Peschel for helpful discussions.
E.J. thanks the Institute for Theoretical Physics
of the University of Fribourg, Switzerland, for its kind hospitality
during the preparation of this manuscript.
S.R.W. acknowledges support from the NSF under Grant No. DMR-98-70930,
and from the University of California through the
Campus Laboratory Collaborations Program.


\end{document}